\documentclass[a4paper]{VisionStyle}
\usepackage{epsfig}

\begin{document}

\title{Abundance gradients and the role of SNe in M87/Virgo}

\author{F.\,Gastaldello\inst{1,2} \and S.\,Molendi\inst{1} } 

\institute{IASF-CNR, via Bassini 15, I-20133 Milano, Italy
\and 
Universit\`a di Milano Bicocca, Dip. di Fisica, P.za della Scienza 3 I-20133 Milano, Italy  }

\maketitle 

\begin{abstract}

We make a detailed measurement of the metal abundance profiles and metal 
abundance ratios of the inner core of M87/Virgo observed by \emph{XMM-Newton} 
during
the PV phase. We use multi temperature models for the inner regions and we
compare the plasma codes APEC and MEKAL. We confirm the strong heavy elements 
gradient previously found by \emph{ASCA} and \emph{BeppoSAX}, but also find a 
significant 
increase in light elements, in particular O. This fact together with the 
constant O/Fe ratio  in the inner 9 arcmin indicates an enhancement 
of contribution in the core of the cluster not only by SNIa but also by SNII.

\end{abstract}

\section{Introduction}
\label{fgastaldello-B3_sec:introduction}

The study of metallicity in clusters of galaxies dates back to 1976, with
the detection of the 7 keV blend line emission due to highly ionized iron as a 
strong feature in the X-ray spectra of the Perseus cluster 
(\cite{fgastaldello-B3:mitchell76}). 
This was a most significant 
observational discovery concerning X-ray clusters, following their 
identification as X-ray sources, because it established without doubt that 
the primary emission mechanism was thermal and that the hot intra-cluster
gas contained a significant portion of processed gas, which had at some point 
been ejected from stars (\cite{fgastaldello-B3:sarazin88}).

The X-ray emitting hot intra-cluster medium (ICM) of clusters of galaxies is
now 
known to contain a large amount of metals: for rich clusters between red-shift 
0.3-0.4 and the present day the observed metallicity is about 1/3 the solar
value (\cite{fgastaldello-B3:mush97}, \cite{fgastaldello-B3:allen98}, \cite{fgastaldello-B3:fuka98}, \cite{fgastaldello-B3:ceca00}, \cite{fgastaldello-B3:ettori01}), 
suggesting that a significant fraction of the ICM has been processed
into stars already at intermediate red-shifts. 
\newline
The total mass of iron is directly proportional to the optical luminosity from
elliptical and lenticular galaxies, this fact strongly arguing for the metals
now in the ICM having been produced by the massive stars of the same stellar
population to which belong the low mass stars now producing the bulk of the
cluster optical light (\cite{fgastaldello-B3:arnaud92}, \cite{fgastaldello-B3:renzini93}).

While the origin of the metals observed in the ICM is clear (they are produced 
by supernovae, in particular SNII and SNIa), less clear is the 
transfer mechanism of
 these metals to the ICM. The main mechanisms that have been 
proposed for the metal enrichment in clusters are: enrichment of gas during the
formation of the proto-cluster (\cite{fgastaldello-B3:kauff98}); 
ram pressure stripping of metal enriched gas from cluster galaxies 
(\cite{fgastaldello-B3:gunn72}, \cite{fgastaldello-B3:toni01}); 
stellar winds AGN- or 
SN-induced in Early-type galaxies (\cite{fgastaldello-B3:matteucci88}, \cite{fgastaldello-B3:renzini97}).

Another controversial subject is the relative contribution to 
the metal enrichment by the two types of supernovae, SNIa and SNII.
\cite*{fgastaldello-B3:mush96} and \cite*{fgastaldello-B3:mush97} showed a 
dominance of SNII ejecta, while other works on \emph{ASCA} data 
(\cite{fgastaldello-B3:ishimaru97}, \cite{fgastaldello-B3:fuka98}, 
\cite{fgastaldello-B3:fino00}, \cite{fgastaldello-B3:dupke00a}), still 
indicating a predominance of SNII enrichment
at large radii in clusters, do not exclude that as much as 50\% 
of the iron in clusters come from SNIa ejecta in the inner part of clusters. 

Spatially resolved abundance measurement in galaxy clusters are of great 
importance because they can be used to measure the precise amounts of metals 
in the ICM and to constrain the origin of metals both spatially and in terms
of the different contributions of the two different type of SNe (since SNIa 
products are iron enriched, while SNII products are rich in $\alpha$ elements 
such as O, Ne, Mg and Si)
as a function of the position in the cluster. The first two satellites 
able to perform spatially resolved spectroscopy, \emph{ASCA} and 
\emph{BeppoSAX} have revealed abundance gradients in cD clusters 
(\cite{fgastaldello-B3:dupke00b}, \cite{fgastaldello-B3:grandi01}),
in particular M87/Virgo (\cite{fgastaldello-B3:matsumoto96}, \cite{fgastaldello-B3:guainazzi00}), and variations in Si/Fe 
within a cluster (\cite{fgastaldello-B3:fino00}) and among clusters 
(\cite{fgastaldello-B3:fuka98}). Given the diversity of metals 
produced by SNIa and SNII, the
variations in Si/Fe suggest that the metals in the ICM have been produced by
a mix of the two types of SNe.
\newline
Moreover with the aid of observational data on many elements, X-ray observation
can discriminate between competing theoretical models for SNIa also 
because the central region of cD clusters, where the SNIa products could 
dominate in the abundance pattern, can be considered as the archives of 
an extremely
 large number of SNIa explosions (\cite{fgastaldello-B3:dupke00b}, \cite{fgastaldello-B3:fino01}). 

The broad bandpass large X-ray sensitivity and good angular resolution of
\emph{XMM-Newton} enables the measurement of abundances and abundance gradients of
elements that have not been accurately studied before. In particular the
O abundance (and O/Fe ratio) is particularly important for estimation of the
SNII contribution since O is expected to be produced mostly by this kind of 
supernova. Limited sensitivities of pre-\emph{XMM-Newton} instruments for O \small{VIII} lines prevent this best measure to separate the enrichment by the two types of supernovae.

M87 is the cD galaxy of the nearest cluster and its high flux and close 
location makes it one of the most ideal target for a detailed study of 
the ICM abundance down to the scale of the kpc.
Throughout this paper, we assume $H_{0}\,=\, 50\, \rm{km\, s^{-1} Mpc^{-1}}$, 
$q_{0} = 0.5$ and at the distance of M87 $1\arcmin$ corresponds to 5 kpc.

\section{Observation and Data Preparation}
\label{fgastaldello-B3_sec:observation}

M87/Virgo was observed with \emph{XMM-Newton} (\cite{fgastaldello-B3:jansen01})
 during the PV 
phase with the MOS detector in Full Frame Mode for an effective exposure
 time of about 39 ks. Details on the observation have been published in 
\cite*{fgastaldello-B3:boh01} and \cite*{fgastaldello-B3:belsole01}.
We have obtained calibrated event files for the MOS1 and MOS2 cameras 
with SASv5.0. Data were manually screened to remove any remaining bright 
pixels or hot column. 
Periods in which the background is increased by soft proton flares have 
been excluded using an intensity filter: we rejected all events accumulated 
when the count rates exceeds 15 cts/100s in the [10 -- 12] keV band for the 
two MOS cameras.

We have accumulated spectra in 10 concentric annular regions centered on the 
emission peak extending our analysis out to 14 arcmin from the emission peak, 
thus exploiting the entire \emph{XMM} field of view. We have removed point 
sources and the substructures which are clearly visible from the X-ray image
(\cite{fgastaldello-B3:belsole01}) except in the innermost region, 
where we have kept
the nucleus and knot A, because on angular scales so small it is not possible 
to exclude completely their emission. We prefer to fit the spectrum of this
region with a model which includes a power law component to fit the two point 
like sources. We include only one power law component due to the similarity
of the two sources spectra (\cite{fgastaldello-B3:boh01}).
The bounding radii are 0$^{\prime}$-0.5$^{\prime}$, 
0.5$^{\prime}$-1$^{\prime}$, 1$^{\prime}$-2$^{\prime}$, 
2$^{\prime}$-3$^{\prime}$, 
3$^{\prime}$-4$^{\prime}$, 
4$^{\prime}$-5$^{\prime}$, 5$^{\prime}$-7$^{\prime}$, 
7$^{\prime}$-9$^{\prime}$, 9$^{\prime}$-11$^{\prime}$ and 
11$^{\prime}$-14$^{\prime}$. 
The analysis of the 4 central 
regions within 3 arcmin was already discussed in 
\cite*{fgastaldello-B3:molegasta01}.

Spectra have been accumulated for MOS1 and MOS2 independently. The Lockman 
Hole observations have been used for the background. Background spectra have 
been accumulated from the same detector regions as the source spectra.

The vignetting correction has been applied to the spectra rather than to the
effective area, as is customary in the analysis of EPIC data 
(\cite{fgastaldello-B3:arnaud01}). Spectral fits were performed in the 
0.5-4.0 keV band. Data below 0.5 keV were excluded to avoid 
residual calibration problems in the MOS response matrices at soft energies.
Data above 4 keV were excluded because of substantial contamination 
of the spectra by hotter gas emitting further out in the cluster, on the
same line of sight.

As discussed in \cite*{fgastaldello-B3:molendi01} there are 
cross-calibration uncertainties 
between
the spectral response of the two EPIC instruments, MOS and PN. In particular 
for what concern the soft energy band (0.5-1.0 keV) fitting six 
extra-galactic spectra for
which no excess absorption is expected, MOS recovered the $\rm{N_{H}}$ 
galactic value, while PN gives smaller $\rm{N_{H}}$ by 
$1-2\times10^{20}\rm{cm^{-2}}$. Thus we think that at the moment the 
MOS results
are more reliable than the PN ones in this energy band, which is crucial 
for the O abundance measure. For this
reason and for the better spectral resolutions of MOS, which is again important
 in deriving the O abundance, we limit our analysis to MOS data.

\section{Spectral modeling and plasma codes}
\label{fgastaldello-B3_sec:spectralmodel}

The X-ray emission in cluster of galaxies originates from the hot gas 
permeating the cluster potential well. The continuum emission is dominated by
thermal bremsstrahlung, which is proportional to the square of the gas density
times the cooling function. From the shape and the normalization of the 
spectrum we derive the gas temperature and density. In addition the X-ray
spectra of clusters of galaxies are rich in emission lines due to K-shell
transitions from O, Ne, Mg, Si, S, Ar and Ca and K- and
L-shell transitions from Fe and Ni, from which we can measure the
relative abundance of a given element. 
A measure of 
the equivalent width of a spectral line is a direct measure of the relative 
abundance of a given element. This comes from the fact that both the 
continuum and line emission are two body processes with the continuum 
emissivity proportional to the electron density times the proton density and 
the line emissivity proportional to the electron density times the density
of a given element. From the definition of equivalent width it is easily 
derived that this quantity is proportional to the ratio between the ion 
and proton densities. 

\begin{figure}[ht]
\vskip -2.5truecm
  \begin{center} 
   \epsfig{file=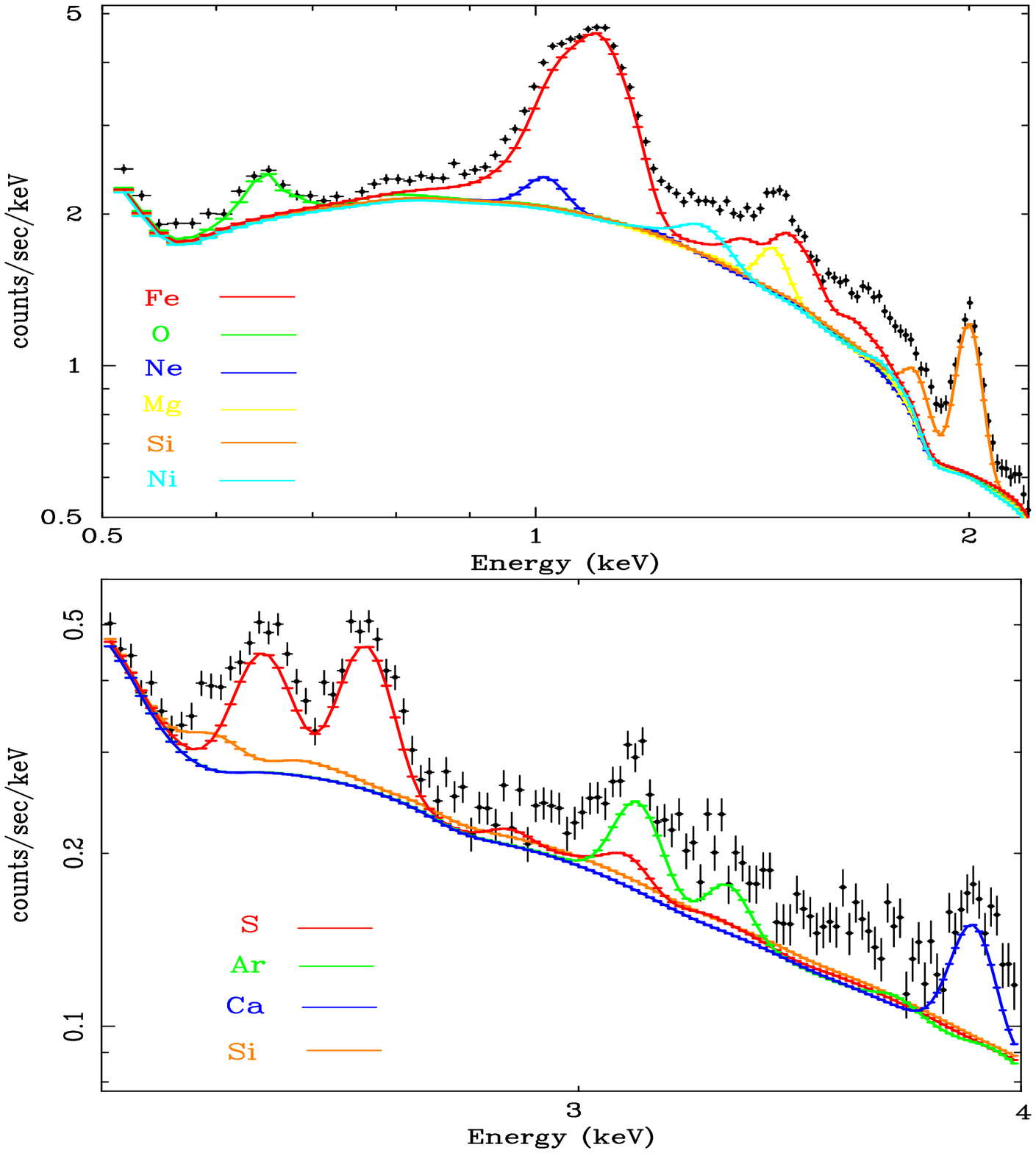, width=9cm}
  \end{center}
\vskip -1.0truecm
\caption{ Data of the 3$^{\prime}$-4$^{\prime}$ bin and lines of the various 
elements, 
obtained by setting all element abundances to zero except the one of interest,
calculated using the MEKAL code.
\newline
{\bf Top Panel}: the lines of O, Fe, Ne, Ni, Mg and Si used for the analysis
 in the 0.5-2 keV energy range.
\newline
{\bf Bottom Panel}: the lines of Si, S, Ar and Ca used for the analysis in the
2-4 keV energy range. }  
\label{fgastaldello-B3_fig:fig1}
\end{figure}

In Figure~\ref{fgastaldello-B3_fig:fig1} we show the data of the 
3$^{\prime}$-4$^{\prime}$ bin together
with the best fitting model calculated using the MEKAL code.
The model has been plotted nine times, each time all element abundances, 
except one, are set to zero. In this way the contribution of the various 
elements to the observed lines and line blends become apparent. In the energy 
band (0.5-4 keV) we have adopted for the spectral fitting, 
the abundance measurements based on
K-lines for all the elements except for Fe and Ni, 
for which the measure is based on L-lines.
The K-lines of O, Si, S, Ar and Ca are well 
isolated 
from other emission features and clearly separated from the continuum emission,
 which are the requirements for a robust measure of the equivalent width of the
lines and consequently of the abundances of these elements. 
The Fe-L lines are known to be problematic, because the atomic physics involved
is more complicated than K-shell transitions (\cite{fgastaldello-B3:liedahl95},
 but from the very good signal of \emph{XMM} spectra and  
from the experience of \emph{ASCA} data  
(\cite{fgastaldello-B3:mush96}, \cite{fgastaldello-B3:hwang97}, 
\cite{fgastaldello-B3:fuka98})
 we can conclude that the Fe-L determination is reliable.
Some of the stronger Fe-L lines due to Fe XXII and Fe XXIV are close to the
K-lines of Ne and Mg, respectively and blending can lead to 
errors in the Ne abundance and, to a smaller extent, to the Mg 
abundance (\cite{fgastaldello-B3:liedahl95}, \cite{fgastaldello-B3:mush96}). 
Also the Ni 
measure is difficult due to the possible confusion of its L-lines with the 
continuum and Fe-L blend.  

The elemental abundances are expressed by the relative values to the solar 
abundances from \cite*{fgastaldello-B3:grevesse98} that updates the commonly 
used values from \cite*{fgastaldello-B3:anders89}

All spectral fitting has been performed using version 11.0.1 of the 
XSPEC package. 

All models discussed below include a multiplicative component
to account for the galactic absorption on the line of sight of M87.
The column density is always fixed at a value of 
$1.8\times 10^{20}\,\rm{cm^{-2}}$, which is derived from 21cm measurements 
(\cite{fgastaldello-B3:lieu96}).
Leaving $\rm{N_{H}}$ to freely vary does not improve the fit and does 
not affect the measure of the oxygen abundance, which could have been 
the more sensitive to the presence of excess absorption. 
The $\rm{N_{H}}$ value 
obtained is consistent within the errors with the 21cm value.

The temperature profile for M87 (\cite{fgastaldello-B3:boh01} shows a small 
gradient for radii larger than $\sim 2$ arcmin and a rapid decrease for 
smaller radii. Moreover, as pointed out in \cite*{fgastaldello-B3:molepizzo01}
  all spectra at radii larger than 2 arcmin are
characterized by being substantially isothermal (although 
the spectra of the regions between 
2 and 7 arcmin are multi temperature spectra with a narrow temperature 
range rather than single temperature spectra), while at radii smaller 
than 2 arcmin
we need models which can reproduce the broad temperature distribution of the
inner regions.

We therefore apply to the central regions (inside 3 arcmin) 
three different spectral models.

A two temperature model 
(vmekal + vmekal in XSPEC and model II in (\cite{fgastaldello-B3:molegasta01}) 
using
the plasma code MEKAL (\cite{fgastaldello-B3:mewe85}, 
\cite{fgastaldello-B3:liedahl95}). 
The metal abundance of each 
element of the second thermal component is bound to be equal to the same 
parameter of the first thermal component. This model is used (e.g. 
\cite*{fgastaldello-B3:makishima01} and refs. therein) as an alternative 
to cooling-flow models in fitting
the central regions of galaxy clusters.

A ``fake multi-phase'' model (vmekal + vmcflow in XSPEC and model III 
in \cite*{fgastaldello-B3:molegasta01}. 
As indicated in recent papers 
(\cite{fgastaldello-B3:molepizzo01}, \cite{fgastaldello-B3:molegasta01}) 
this model is used to
describe a scenario different from a multi-phase gas, for which it was 
written for: the gas is all at one temperature and the multi-phase 
appearance of the spectrum comes from projection of emission from many 
different physical radii. A more correct description will be given by a real
deprojection of the spectrum.
  
The third model is the analogue of the vmekal two temperature model using
the plasma code APEC (\cite{fgastaldello-B3:smith01}). 

We can't adopt an APEC analogue of the fake multi-phase model because the 
cooling flow model calculating its emission using APEC is still 
under development. Given the substantial agreement between 2T and fake 
multi-phase model (\cite{fgastaldello-B3:molegasta01}), we can regard the 
2T APEC 
results as indicative also for a fake multi-phase model.

For the spectrum accumulated in the innermost region we included also a power
law component to model the emission of the nucleus and of knot A.

\begin{figure}[hb!]
\vskip -0.5truecm
  \begin{center} 
   \epsfig{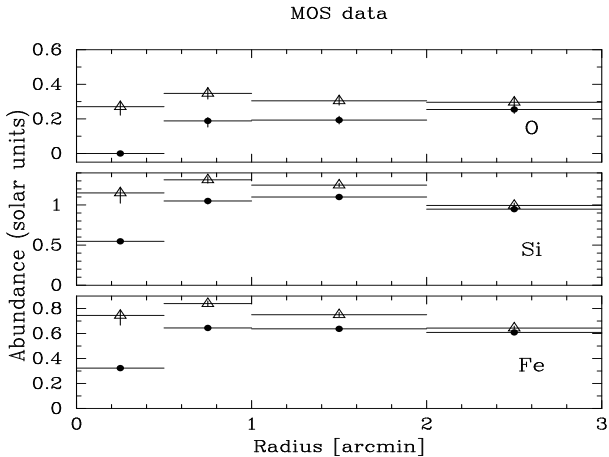}
  \end{center}
\caption{ Abundance profiles for O, Si and Fe in the inner 3 arcmin. 
Full circles indicate the measurements with a single temperature model, while 
empty triangles those with a two temperature model. Solar units are the old ones by Anders \& Grevess (1989). } 
\label{fgastaldello-B3_fig:fig2}
\end{figure}

For what concern the measure of metal abundances, it is crucial to use multi 
temperature models in the inner bins, where the evidence of multi temperature 
gas is stronger. Infact single temperature models substantially underestimate 
the metal abundance in multi temperature spectra, as already observed in ASCA 
spectra of galaxies and galaxy groups 
(Fe bias: \cite{fgastaldello-B3:buote00}). In M87 the drastic abundance 
decrement found in the previous analysis (\cite{fgastaldello-B3:boh01}) 
follows from the application of the oversimplified single temperature model
(\cite{fgastaldello-B3:molegasta01}), as shown in 
Figure~\ref{fgastaldello-B3_fig:fig2}.

For the outer regions (from 3 arcmin outwards) we apply single temperature
models: vmekal using the MEKAL code 
and vapec using the APEC code.

We use the two different plasma codes with the aim of cross-checking their
results. 
As pointed out by the authors of the new code APEC, cross-checking is very 
important, since each plasma emission code requires choosing from a large 
overlapping but incomplete set of atomic data and the results obtained by using
independent models allows critical comparison and evaluation of errors in
the code and in the atomic database.

\section{Results}
\label{fgastaldello-B3_sec:results}
\subsection{Abundance profiles}
\label{fgastaldello-B3_sec:abprofiles}

\begin{figure}[ht!]
\vskip -1.0truecm
\hskip -2.0truecm 
   \epsfig{file=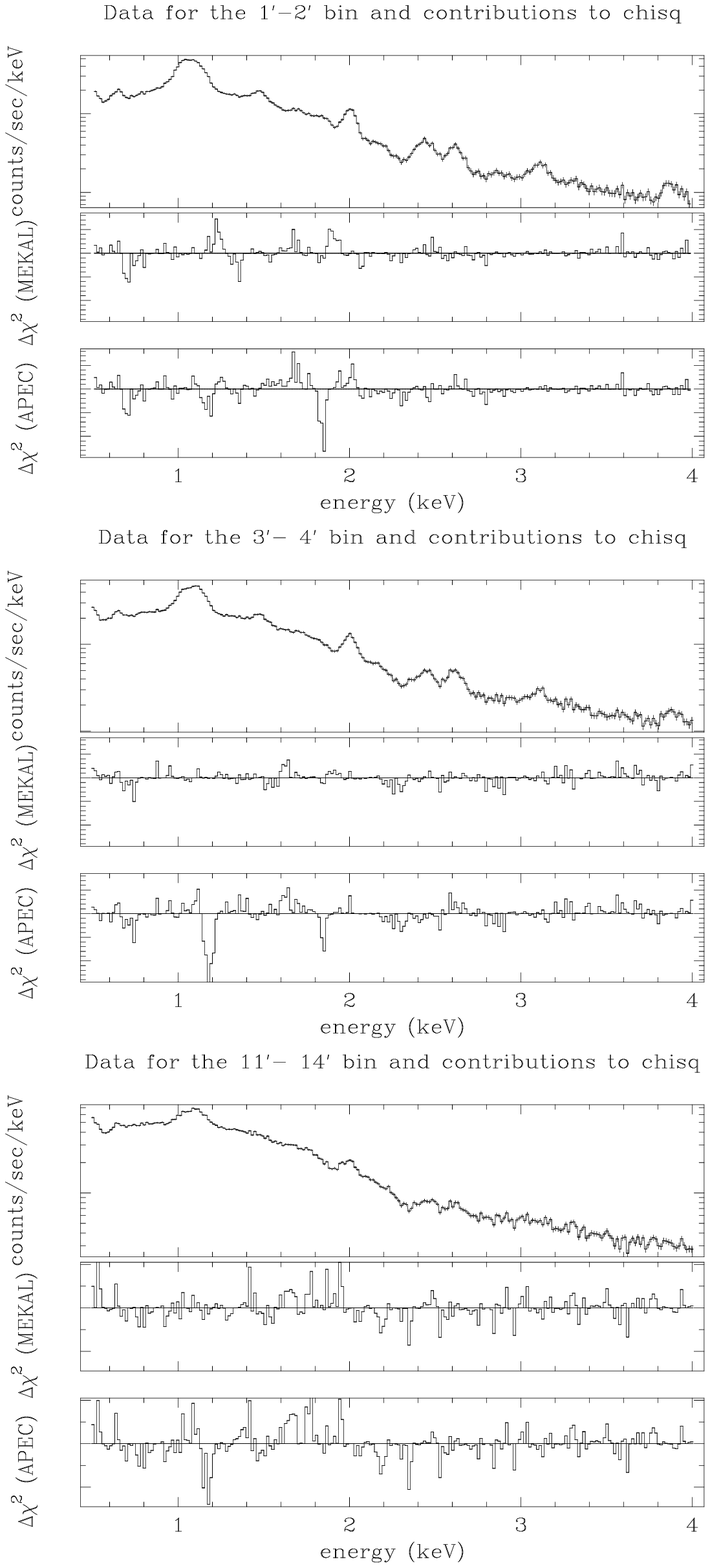, height=12cm,width=12cm}
\vskip -0.5truecm
\caption{ {\bf Top Panel}: data for the bin 
1$^{\prime}$-2$^{\prime}$ and contributions to $\chi^{2}$ for a 2T model with MEKAL code, the $\chi^{2}/d.o.f.$ of the fit is 381/217, and for a 2T model with APEC code, the $\chi^{2}/d.o.f.$ of the fit is 474/218. Contributions to $\chi^{2}$ are on the same scale  for direct comparison.
\newline
{\bf Middle Panel}: data for the bin 
3$^{\prime}$-4$^{\prime}$ and contributions to $\chi^{2}$ for a 1T model with MEKAL code, the $\chi^{2}/d.o.f.$ of the fit is 326/221, and for a 
1T model with APEC code, the $\chi^{2}/d.o.f.$ of the fit is 
546/222. Contributions to $\chi^{2}$ are on the same scale for direct 
comparison. 
\newline
{\bf Bottom Panel}: data for the bin 
11$^{\prime}$-14$^{\prime}$ and contributions to $\chi^{2}$ for a 1T model with MEKAL code, the $\chi^{2}/d.o.f.$ of the fit is 729/221, and for a 
1T model with APEC code, the $\chi^{2}/d.o.f.$ of the fit is 
886/222. Contributions to $\chi^{2}$ are on the same scale for direct 
comparison. }  
\label{fgastaldello-B3_fig:fig3}
\end{figure}

The abundance measurements obtained using the two different plasma codes agree
for what concerns Fe, Ar and Ca; they are somewhat 
different for what concerns O, S and Ni and in complete 
disagreement for Mg and Ne. The temperature profile obtained 
with the two codes are somewhat different in the inner regions (where we 
compare the range of temperature obtained by multi-temperature models), while 
there is substantially agreement in the outer isothermal bins.

The models using the APEC code give a systematically worse description than
the ones using MEKAL code; we show some examples in Fig.~\ref{fgastaldello-B3_fig:fig3}. In the external bins the differences in the fit between the two codes are due
to APEC over-prediction of the flux of Fe-L lines from high ionization states, 
considering the fact that the temperature obtained by the two codes are 
nearly coincident.
For the inner regions the differences between the two codes are further 
complicated by the different temperature range they find for the best fit.
In general, where the temperature structure is very similar, as in the 
innermost bin, the
difference is as in the outer bins in the high energy part of the Fe-L blend,
 while where the temperature structure is different, as in the 
1$^{\prime}$-2$^{\prime}$ shown in Figure~\ref{fgastaldello-B3_fig:fig3}, the differences between the two 
codes are primary due to different estimates of the flux of He-like Si-K line.

\begin{figure}[h!]
  \begin{center} 
   \epsfig{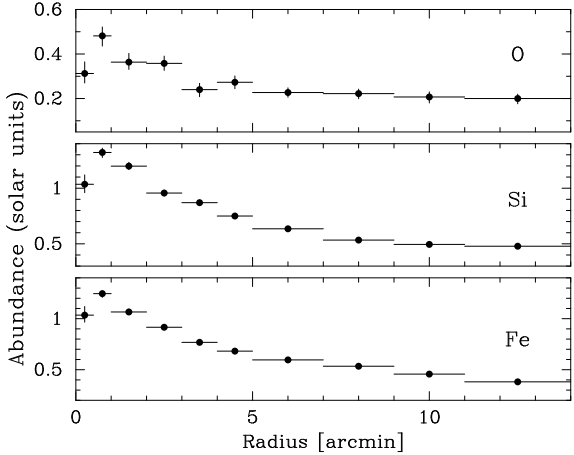}
  \end{center}
\caption{ MOS abundance profiles for O, Si and Fe, obtained with a 2T model 
using MEKAL code for the inner 3 arcmin and with a 1T model using MEKAL code 
for the outer bins. Uncertainties are at the 68\% 
level for one interesting parameter($\Delta \chi^{2}\,=\,1$).  }  
\label{fgastaldello-B3_fig:fig4}
\end{figure}

\begin{figure}[h!]
  \begin{center} 
   \epsfig{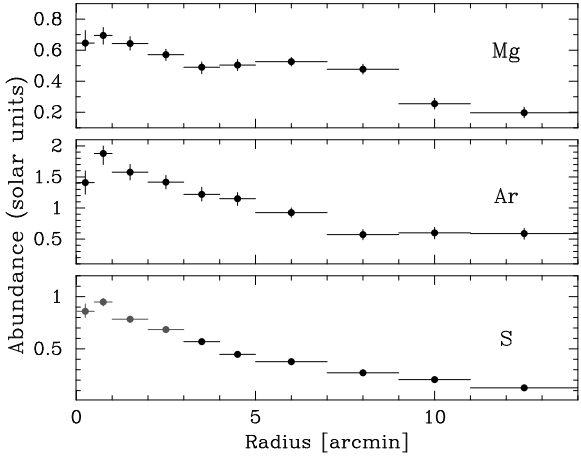}
   \end{center}
\caption{ Same as Fig.~\ref{fgastaldello-B3_fig:fig4} but for Mg, Ar and S. }  
\label{fgastaldello-B3_fig:fig5}
\end{figure}

\begin{figure}[h!]
  \begin{center} 
   \epsfig{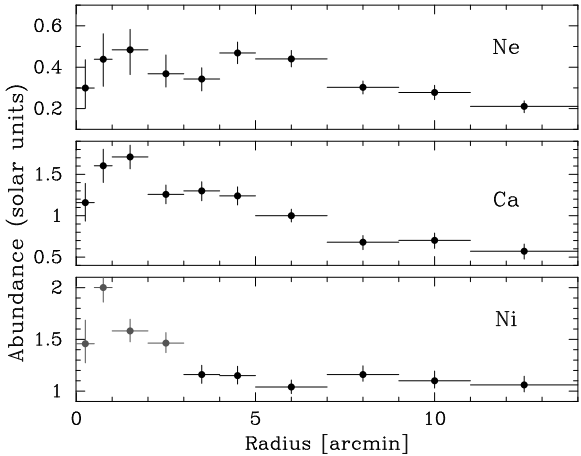}
  \end{center}
\caption{Same as Fig.~\ref{fgastaldello-B3_fig:fig4} but for Ne, Ca and Ni. }  
\label{fgastaldello-B3_fig:fig6}
\end{figure}

We therefore choose as our best abundance profiles those obtained with 
a 2T vmekal fit for the central regions and with a 1T vmekal fit for the outer
regions.
We report the abundance profiles obtained in this way
in Fig~\ref{fgastaldello-B3_fig:fig4} for O, Si, Fe,
in Fig~\ref{fgastaldello-B3_fig:fig5} for 
Mg, Ar, S and in Fig~\ref{fgastaldello-B3_fig:fig6} for 
Ne, Ca, Ni.
Abundance gradients are clearly evident for Fe, Si, S,
Ar and Ca; a statistically significant enhancement 
is evident in the central regions 
for O, Mg and Ni, while only Ne 
is substantially flat.

\subsection{Abundance ratios}
\label{fgastaldello-B3_sec:abratios}

\begin{figure}[ht!]
  \begin{center} 
   \epsfig{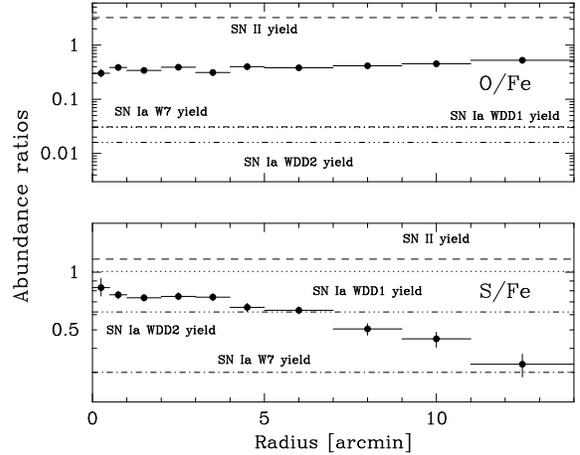}
  \end{center}
\caption{ MOS abundance ratio profiles for O/Fe and S/Fe. Also showed are the 
abundance ratios predicted by SNe models taken by 
Nomoto et al. 1997: the dashed
line refers to the SNII model, the dash-dotted line to the W7 SNIa model, the
dotted line to the WDD1 SNIa model and the three dotted-dashed line to the 
WDD2 SNIa model. }  
\label{fgastaldello-B3_fig:fig7}
\end{figure}

From the abundance measurements we obtain the abundance ratios between all
the elements relative to Fe, normalized to the
solar value. As an example we show the two more significant abundance ratios, 
O/Fe and S/Fe, in Fig.~\ref{fgastaldello-B3_fig:fig7}, together with the 
abundance ratios obtained 
by models of supernovae taken by \cite*{fgastaldello-B3:nomoto97} and 
rescaled to the 
solar abundances reported in \cite*{fgastaldello-B3:grevesse98}.

An important results is the constance in the inner 9 arcmin of the ratio
O/Fe (fitting the data with a constant gives $\chi^{2}=6.9$ for 7 d.of. and 
adding a linear component does not improve the fit, $\chi^{2}=5.4$ for 6 d.o.f)
which, together with the enhancement in the O abundance in the inner regions 
points towards an increase in contribution by SNII, since O is basically 
produced only by this kind of supernovae.

The ratio S/Fe, and in general M87 data, suggests an agreement with delayed 
detonation models (in particular for the inner bins), as stressed by 
\cite*{fgastaldello-B3:fino01}. If we consider the set of theoretical 
values for
SNII and W7 SNIa models the behavior of these ratio would indicate an 
increasing contribution by SNIa going \emph{outward} to the center. 
We recover the
correct behavior if we choose the WDD1 yield and we reduce the S SNII yield of
\cite*{fgastaldello-B3:nomoto97} by a factor of two to three, as was already 
indicated by ASCA data (\cite{fgastaldello-B3:dupke00a}).

\section{Discussion}
\label{fgastaldello-B3_sec:discussion}

The model emerging from the \emph{ASCA} and \emph{BeppoSAX} data for the 
explanation of 
abundance gradients in galaxy clusters was that of a homogeneous enrichment by
SNII, the main source of $\alpha$ elements, maybe in the form of strong 
galactic winds in the proto-cluster phase and
the central increase in the heavy element distribution due to an enhanced 
contribution by SNIa, strongly related to the presence of a cD galaxy 
(\cite{fgastaldello-B3:fuka98}, \cite{fgastaldello-B3:dupke00b}, \cite{fgastaldello-B3:fino00}, \cite{fgastaldello-B3:grandi01}, \cite{fgastaldello-B3:makishima01}).
As a textbook example we can consider the case of A496 observed by \emph{XMM-Newton} 
(\cite{fgastaldello-B3:tamura01}). 
The O-Ne-Mg abundance is radially constant over the 
cluster, while the excess of heavy elements as Fe, Ar, Ca and Ni in the core is
consistent with the assumption that the metal excess is solely produced by
SNIa in the cD galaxy. The crucial ratio for the discrimination of the 
enrichment by the two types of supernovae, O/Fe, is then decreasing towards 
the center.

The XMM results for M87/Virgo complicates this picture. Together with the 
improved accuracy in the measure of the heavy elements gradients, showing
the increased contribution by SNIa, the statistically significant enhancement
of  $\alpha$ elements O and Mg and the constant O/Fe ratio point
toward an increase in 
contribution also of SNII. 
A mechanism which 
enhances the contribution of SNII over the homogeneous contribution all over 
the cluster must have been at work in the center of the cluster.
Although there is little or no evidence of current star formation 
in the core of M87, the O excess could  be related to a recent past episode of
 star formation triggered
by the passage of the radio jet, as we see in cD galaxies with a radio source 
(A1795 cD: \cite{fgastaldello-B3:vanbreugel84} and A2597 cD: \cite{fgastaldello-B3:koekemoer99}), nearby (Cen A: \cite{fgastaldello-B3:graham98}) and
distant radio galaxies (\cite{fgastaldello-B3:vanbreugel85}, \cite{fgastaldello-B3:vanbreugel93}, \cite{fgastaldello-B3:bicknell00}) 
(for a comprehensive discussion see \cite{fgastaldello-B3:mcnamara99}).
\newline 
To put the above idea quantitatively, waiting for a true deprojection of our 
data, we use previous ROSAT estimate of the deprojected electron density 
in the center of 
the Virgo cluster (\cite{fgastaldello-B3:nulsen95}) to calculate the excess 
mass of oxygen. To 
estimate the excess abundance we fit the inner bins, where we see the 
stronger increase in the O abundance, with a constant obtaining an abundance
of 0.32, while for the outer bins we obtain an abundance of 0.21 so the excess
is 0.11. Then we estimate the oxygen mass, $M_{\rm{O}}$ to be $M_{\rm{O}} = A_{\rm{O}}\,y_{\rm{O},\odot}\,Z_{\rm{O}}^{excess}\,M_{H}$, where $M_{H} = 0.82\,n_{e}\,\frac{4\pi}{3}\,(R_{out}-R_{in})^{3}$, $R_{in}$ and $R_{out}$ being the bounding radii in kpc of the bins,  $A_{\rm{O}}= 16$ and $y_{\rm{O},\odot} = 6.76 \times 10^{-5}$ from \cite*{fgastaldello-B3:grevesse98}. We obtain a rough estimate of $10^{7}\,\rm{M_{\odot}}$ which, assuming 
that 
about $80\,\rm{M_{\odot}}$ of star formation are required to generate a SNII 
(\cite{fgastaldello-B3:thomas90}) and Nomoto SNII oxygen yield, requires a cumulative star 
formation of $4-5 \times 10^{7}\,\rm{M_{\odot}}$. This star formation is in 
agreement with a burst mode of star formation ($\sim 10^{7}$ yr) at 
rates of 
$\sim 10-40\,\rm{M_{\odot}\,yr^{-1}}$, as it is observed in the CD 
galaxies of A1795 and A2597 (\cite{fgastaldello-B3:mcnamara99}).

\begin{acknowledgements}

 This work is based on observations obtained with \emph{XMM-Newton}, an ESA science mission with instruments and contributions directly funded by ESA Member States and the USA (NASA).

\end{acknowledgements}


\begin{thebibliography}{}

\bibitem[\protect\astroncite{Allen \& Fabian}{1998}]{fgastaldello-B3:allen98} Allen, S. W. \& Fabian, A. C. 1998, MNRAS, 297, L63

\bibitem[\protect\astroncite{Anders \& Grevesse}{1989}]{fgastaldello-B3:anders89} Anders, E. \& Grevesse, N. 1989, Geochimica et Cosmochimica Acta, 53, 197

\bibitem[\protect\astroncite{Arnaud et al.}{1992}]{fgastaldello-B3:arnaud92} Arnaud, M., Rothenflug, R., Boulade, O., Vigroux, L., Vangioni-Flam, E., 1992, A\&A, 254, 49

\bibitem[\protect\astroncite{Arnaud et al.}{2001}]{fgastaldello-B3:arnaud01} Arnaud, M., Neumann, D. M., Aghanim, N., Gastaud, R., Majerowicz, S., Hughes, J. P. 2001, A\&A, 365, L80

\bibitem[\protect\astroncite{Belsole et al.}{2001}]{fgastaldello-B3:belsole01} Belsole, E., Sauvageot, J. L., B\"{o}hringer, H., Worral, D. M., Matsushita, K., Mushotzky, R. F., Sakelliou, I., Molendi, S., Ehle, M., Kennea, J., Stewart, G., Vestrand, W. T. 2001, A\&A, 365, L188

\bibitem[\protect\astroncite{Bicknell et al.}{2000}]{fgastaldello-B3:bicknell00} Bicknell, G., Sutherland, R., van Breugel, W.J.M., Dopita, M.A., 
Dey, A, Miley, G.K. 2000, ApJ, 540, 678


\bibitem[\protect\astroncite{B\"{o}hringer et al.}{2001}]{fgastaldello-B3:boh01} B\"{o}hringer, H., Belsole, E., Kennea, J., Matsushita, K., Molendi, S., Worral, D. M., Mushotzky, R. F., Ehle, M., Guainazzi, M., Sakelliou, I., Stewart, G., Vestrand, W. T., Dos Santos, S. 2001, A\&A, 365 L181

\bibitem[\protect\astroncite{Buote}{2000}]{fgastaldello-B3:buote00} Buote, D.A., 2000, ApJ, 539, 172

\bibitem[\protect\astroncite{De Grandi \& Molendi}{2001}]{fgastaldello-B3:grandi01} De Grandi, S. \& Molendi, S. 2001, ApJ, 551, 153

\bibitem[\protect\astroncite{Della Ceca et al}{2000}]{fgastaldello-B3:ceca00} Della Ceca, R., Scaramella, R., Gioia, I. M., Rosati, P., Fiore, F., Squires, G. 2000, A\&A, 353, 498

\bibitem[\protect\astroncite{Dupke \& White}{2000a}]{fgastaldello-B3:dupke00a} Dupke, R. A. \& White R. 2000a, ApJ, 528, 139

\bibitem[\protect\astroncite{Dupke \& White}{2000b}]{fgastaldello-B3:dupke00b} Dupke, R. A. \& White R. 2000b, ApJ, 537, 123

\bibitem[\protect\astroncite{Ettori et al.}{2001}]{fgastaldello-B3:ettori01} Ettori, S., Allen, S. W., Fabian, A. C. 2001, MNRAS, 322, 187 

\bibitem[\protect\astroncite{Finoguenov et al}{2000}]{fgastaldello-B3:fino00} Finoguenov, A., David, L. P., Ponman, T. J. 2000, ApJ, 544, 188

\bibitem[\protect\astroncite{Finoguenov et al}{2001}]{fgastaldello-B3:fino01} Finoguenov, A., Matsushita, K., B\"ohringer, H., Ikebe, Y., Arnaud, M. 2001,  A\&A in press, (astro-ph/0110516)

\bibitem[\protect\astroncite{Fukazawa et al.}{1998}]{fgastaldello-B3:fuka98} Fukazawa, Y., Makishima, K., Tamura, T., Ezawa, H., Xu, H., Ikebe, Y., kikuchi, K., Ohashi, T. 1998, PASJ, 50, 187

\bibitem[\protect\astroncite{Graham}{1998}]{fgastaldello-B3:graham98} Graham, J.A., 1998, ApJ, 502, 245

\bibitem[\protect\astroncite{Gibson et al.}{1997}]{fgastaldello-B3:gibson97} Gibson, B. K., Lowenstein, M., Mushotzky, R. F. 1997, MNRAS, 290, 623

\bibitem[\protect\astroncite{Grevesse \& Sauval}{1998}]{fgastaldello-B3:grevesse98} Grevesse, N. \& Sauval, A. J., 1998, Space Science Reviews, 85, 161

\bibitem[\protect\astroncite{Guainazzi \& Molendi}{2000}]{fgastaldello-B3:guainazzi00} Guainazzi, M. \& Molendi, S. 2000, A\&A, 351, L19 

\bibitem[\protect\astroncite{Gunn \& Gott}{1972}]{fgastaldello-B3:gunn72} Gunn, J. E. \& Gott, J. R. 1972, ApJ, 176, 1

\bibitem[\protect\astroncite{Hwang et al.}{1997}]{fgastaldello-B3:hwang97} Hwang, U., Mushotzky R., Loewenstein, M., Markert, T. H., Fukazawa, Y., Matsumoto, H. 1997, ApJ, 476, 560

\bibitem[\protect\astroncite{Kauffman \& Charlot}{1998}]{fgastaldello-B3:kauff98} Kauffmann, G. \& Charlot, S. 1998, MNRAS, 294, 705

\bibitem[\protect\astroncite{Koekemoer et al.}{1999}]{fgastaldello-B3:koekemoer99} Koekemoer, A.M., O'Dea, C.P., Sarazin, C.L., McNamara, B.R., Donahue, M., Voit, G.M., Baum, S.A., Gallimore, J.F., ApJ, 525, 621

\bibitem[\protect\astroncite{Jansen et al.}{2001}]{fgastaldello-B3:jansen01} Jansen, F., Lumb, D., Altieri, B., Clavel, J., Ehle, M., Erd, C., Gabriel, C., Guainazzi, M., Gondoin, P., Much, R., Munoz, R., Santos, M.,Schartel, N., Texier, D., Vacanti, G. 2001, A\&A, 365, L1

\bibitem[\protect\astroncite{Ishimaru \& Arimoto}{1997}]{fgastaldello-B3:ishimaru97} Ishimaru, Y. \& Arimoto, N. 1997, PASJ, 49,1 

\bibitem[\protect\astroncite{Liedahl et al.}{1995}]{fgastaldello-B3:liedahl95} Liedahl, D. A., Osterheld A. L., Goldstein, W. H. 1995, ApJ, 438, L115 

\bibitem[\protect\astroncite{Lieu et al.}{1996}]{fgastaldello-B3:lieu96} Lieu, R., Mittaz, J. P. D., Bowyer, S., Lockman, F. J., Hwang,, C. Y., Schmitt, J. H. M. M. 1996, ApJ, 458, L5

\bibitem[\protect\astroncite{Makishima et al.}{2001}]{fgastaldello-B3:makishima01} Makishima, K., Ezawa, H., Fukazawa, Y., Honda, H., Ikebe, Y., Kamae, T., Kikuchi, K., Matsushita, K., Nakazawa, K., Ohashi, T., Takahashi, T., Tamura, T., Xu, H. 2001, PASJ, 53, 401

\bibitem[\protect\astroncite{Matsumoto et al.}{1996}]{fgastaldello-B3:matsumoto96} Matsumoto, H., Koyama, K., Awaki, H., Tomida, H., Tsuru, T., Mushotzky, R., Hatsukade, I. 1996, PASJ, 48, 201

\bibitem[\protect\astroncite{Matteucci \& Vettolani}{1988}]{fgastaldello-B3:matteucci88} Matteucci, F. \& Vettolani, G. 1988, A\&A, 202, 21

\bibitem[\protect\astroncite{McNamara}{1999}]{fgastaldello-B3:mcnamara99} McNamara 1999, presented at ``Life cycles of Radio Galaxies'', July 15-17, 1999, STScI, Baltimore (astro-ph/9911129)

\bibitem[\protect\astroncite{Mewe et al.}{1985}]{fgastaldello-B3:mewe85} Mewe, R., Gronenschild, E. H. B. M., van den Oord, G. H. J., 1985, A\&AS, 62, 197

\bibitem[\protect\astroncite{Mitchell et al.}{1976}]{fgastaldello-B3:mitchell76} Mitchell, R.J., Culhane, R.J., Davison, P.J., Ives, J.C., 1976, MNRAS, 176, 29p

\bibitem[\protect\astroncite{Molendi}{2001}]{fgastaldello-B3:molendi01} Molendi, S., 2001, {\it{Report on MOS PN cross-calibration presented at Leicester EPIC calibration meeting held in June 2001}}


\bibitem[\protect\astroncite{Molendi \& Gastaldello}{2001}]{fgastaldello-B3:molegasta01} Molendi, S. \& Gastaldello, F. 2001, A\&A, 375, L14

\bibitem[\protect\astroncite{Molendi \& Pizzolato}{2001}]{fgastaldello-B3:molepizzo01} Molendi, S. \& Pizzolato, F. 2001, ApJ, 560, 194

\bibitem[\protect\astroncite{Mushotzky et al.}{1996}]{fgastaldello-B3:mush96} Mushotzky, R., Loewenstein, M., Arnaud, K. A., Tamura, T., Fukazawa, Y., Matsushita, K., Kikuchi, K., Hatsukade, I. 1996, ApJ, 466, 686

\bibitem[\protect\astroncite{Mushotzky \& Lowenstein}{1997}]{fgastaldello-B3:mush97} Mushotzky, R. \& Lowenstein, M. 1997, ApJ, 481, L63

\bibitem[\protect\astroncite{Nomoto et al.}{1997}]{fgastaldello-B3:nomoto97} Nomoto, K., Iwamoto, K., Nakasato, N., Thielemann, F. K., Brachwitz, F., Tsujimoto, T., Kubo, Y., Kishimoto, N. 1997, Nucl.Phys. A, 621, 467

\bibitem[\protect\astroncite{Nulsen \& B\"{o}hringer}{1995}]{fgastaldello-B3:nulsen95} Nulsen, P.E.J. \& B\"{o}hringer, H., 1995, MNRAS, 274, 1093

\bibitem[\protect\astroncite{Renzini et al.}{1993}]{fgastaldello-B3:renzini93} Renzini, A., Ciotti, L., D'Ercole, A., Pellegrini, S., 1993, ApJ, 419, 52

\bibitem[\protect\astroncite{Renzini}{1997}]{fgastaldello-B3:renzini97} Renzini, A., 1997, ApJ, 488, 35

\bibitem[\protect\astroncite{Sarazin}{1988}]{fgastaldello-B3:sarazin88} Sarazin, C.L., X-ray emission from clusters of galaxies. Cambridge Univ. Press, Cambridge

\bibitem[\protect\astroncite{Smith et al.}{2001}]{fgastaldello-B3:smith01} Smith, R. K., Brickhouse, N. S., Liedahl, D. A., Raymond, J. C. 2001, ApJ, 556, L91

\bibitem[\protect\astroncite{Tamura et al.}{2001}]{fgastaldello-B3:tamura01} Tamura, T., Bleeker, A.M., Kaastra, J.S., Ferrigno, C., Molendi, S. 2001, A\&A, 379, 107

\bibitem[\protect\astroncite{Thomas \& Fabian}{1990}]{fgastaldello-B3:thomas90} Thomas, P.A. \& Fabian, A.C., 1990, MNRAS, 246, 156

\bibitem[\protect\astroncite{Toniazzo \& Schindler}{2001}]{fgastaldello-B3:toni01} Toniazzo, T. \& Schindler, S. 2001, MNRAS, 325, 509

\bibitem[\protect\astroncite{van Breugel et al.}{1984}]{fgastaldello-B3:vanbreugel84} van Breugel, W.J.M., Heckman, T., Miley, G. 1984, ApJ, 276, 79

\bibitem[\protect\astroncite{van Breugel et al.}{1985}]{fgastaldello-B3:vanbreugel85} van Breugel, W.J.M., Filippenko, A.V., Heckman, T., Miley, G. 1985, ApJ, 293, 83

\bibitem[\protect\astroncite{van Breugel \& Dey}{1993}]{fgastaldello-B3:vanbreugel93} van Breugel, W.J.M. \& Dey, A. 1993, ApJ, 414, 563



\end{thebibliography}
\end{document}